\documentclass[a4paper,11pt]{article}
\pdfoutput=1 

\usepackage{jheppub} 

\usepackage[T1]{fontenc} 
\usepackage{float}
\usepackage{epstopdf}
\usepackage{amsmath}
\makeatletter
\gdef\@fpheader{}
\makeatother

\title{\boldmath Topological black holes of Einsteinian cubic gravity and Born-Infeld-type electrodynamics}

\author[1]{Askar Ali}

\affiliation[1]{Department of Sciences and Humanities, National University of Computer and Emerging Sciences, Peshawar 25000, Pakistan}

\emailAdd{askarali@math.qau.edu.pk}

\abstract{In this paper, we examine the new black holes of Einsteinian cubic gravity within the framework of Born-Infeld-type electrodynamics. Initially, we establish the differential equations of motion that characterize an innovative collection of charged black hole solutions in Einsteinian cubic theory. Next, we compute the thermodynamic quantities of these black holes for the case of vanishing bare cosmological constant. We consider this situation because the model is unitary only for the asymptotically flat solutions that allow horizons with spherical topology. We also present that the first law of thermodynamics is fulfilled for these objects. Finally, we examine how the nonlinearity of electric field, electric charge and cubic coupling parameter can impact the local thermal stability of our simulated black holes in both canonical and grand canonical ensembles.
\vspace{80 mm}
}

\notoc 

\begin{document}
\maketitle
\flushbottom 


\section{Introduction}
\label{sec:intro}
All the inquiries were adequately passed by Einstein's theory of gravity (ETG). Just over a century after Einstein's theory, the most significant was actually the revelation of gravitational wave \cite{1ar}. Even with all these breakthroughs, it is inevitably going to adjust the model of ETG whenever spacetime curvature is exceedingly enormous, say, close to a central singularity. Arguably the most logical adjustment is to consider the higher-order curvature terms in the gravitational Lagrangian. The widely recognized higher-order Lovelock terms offer this variety of amendment, which adhere to the restriction of preliminary model of ETG \cite{2ar,3ar}. Nonetheless, this inclusion does not contribute to dynamical equations in four dimensions. Einsteinian cubic gravity (ECG), a cubic order curvature model featuring significant effects in four dimensions, was merely put forward \cite{4ar}. This theoretical framework has earned a considerable amount of interest \cite{5ar,6ar,7ar,8ar,9ar,10ar,11ar,12ar,13ar,14ar,15ar,16ar,17ar,18ar,19ar}. Numerous of the limitations of Lovelock models are recognized in ECG as well. For instance, on a maximally symmetric background, it simply propagates a transverse and massless graviton. Additionally, the relative coefficients of the various curvature invariants within this framework are the same across all dimensions. The dynamical impact in four dimensions alongside additional aspects allow this model desirable and significant. These physical aspects enable us to figure out the impacts of higher-order curvature adjustments on $(2+1)$-dimensional holographic duals of gravity theory solutions. 

The solutions in ECG were probed from several perspectives. In Ref. \cite{5ar}, the perturbative five-dimensional black hole solution within this model has been utilized to estimate the holographic entanglement R\'{e}nyi entropy in the context of dual field theory. The earliest illustrations of black hole solutions in ECG were reported in Ref. \cite{6ar}. The static and spherically symmetric (SSS) analogues of four-dimensional charged and neutral black hole solutions in ECG along with their physical features were investigated in Ref. \cite{7ar}. The most generic model of gravity up to third order in curvature called generalized quasi-topological gravity (GQTG) is outlined in Ref. \cite{8ar}. It needs to be emphasized that a single field equation precisely illustrates the SSS vacuum solutions of GQTG. In the backdrop of GQTG, the ECG together with Lovelock and quasi-topological theories were subsequently acquired in four dimensions as its exceptional scenarios. Additionally, extensive results concerning SSS solutions of generic higher-derivative theories, especially GQTG, have been reported \cite{9ar}. It was also previously demonstrated that, within an appropriate mass, the four-dimensional black hole solutions belonging to an infinite class of ghost-free higher-order theories remain unconditionally stable \cite{10ar}. By adopting the continuous fraction estimation, certain intriguing characteristics of black holes of ECG, notably the black hole's shadow and circular motion of massive particles close to a black hole, were addressed \cite{11ar}. Multiple aspects of a $(2+1)$-dimensional non-supersymmetric conformal field theories that may be equivalent to a holographic dual relative to ECG in four dimensions were reported in Ref. \cite{12ar}. Furthermore, the Euclidean AdS-Taub-NUT and bolt solutions with different base topologies in four and six dimensions were determined in ECG and GQTG, accordingly \cite{13ar}.          

The intent of this inquiry is to address the four-dimensional topological black holes of ECG induced from the formulation of nonlinear electrodynamics (NLED). Numerous types of NLED models were checked out in an attempt to design fresh types of solutions. Every model of NLED has beneficial characteristics of its own \cite{7sw,7saw,7sbw,7scw,7sdw,7Ahendi,7Bhendi,7sew,7sfw,7sgw}. In contrasting with Maxwell's formalism, the NLED models recommend theoretical significance for navigating with the finite attitude of electric field and potentials \cite{8sw,9sw,10sw,11sw,12sw}. One out of several approaches for the formulation of NLED, the Born-Infeld model \cite{7sw}, has been developed to deduce a finite self-energy of electrons. The underlying solution of ETG based on this presumption of matter source was figured out in Ref. \cite{13sw}. Later, more progressively charged black hole solutions of ETG coupled to NLED were thoroughly examined \cite{14sw,15sw,16sw,17sw,18sw,19sw,20sw,21sw,22sw,23sw,24sw,25sw,26sw,27sw}. Meanwhile, innovative black hole solutions in Lovelock theory sourced by NLED were uncovered \cite{29sw,30sw,31sw,32sw,33sw,34sw,34saw,35sw,36sw,38sw}. It was additionally examined how nonlinear electromagnetic sources affect the physical characteristics of exotic Lovelock black holes \cite{40sw,40askar,41askar}. In furtherance, the rotating black branes of ETG coupled to NLED were analyzed \cite{47W}. Plus, the procedure used to construct these solutions in Gauss-Bonnet theory was established \cite{48W}. The thermodynamic attributes of these structures in Lovelock theory with Maxwellian \cite{49W,49H} and NLED sources \cite{50W,51W,52W} were additionally discussed. Subsequently, rotating black branes of quasi-topological gravity supported by NLED have been further sought out \cite{53W,54W}. The physical features of new topological black holes in ECG linked to the Born-Infeld electrodynamics were addressed in Ref. \cite{1cubic}. Presently, we are planning to put an emphasis on analyzing charged black holes in ECG arising from two distinct models of electromagnetism, namely the exponential (EN) and logarithmic (LN) forms of NLED. These two models have been developed for several purposes, but the main justification for adopting these approaches is that they might be estimated via an analysis of the loop corrections \cite{55A,55B,55C,55D}. Note that even though the original Born-Infeld model and the other Born-Infeld-type formulations such as EN and LN models exhibit a few features in common, yet they distinguish themselves in substantial ways \cite{7sbw,7scw,7sdw,7Ahendi,7Bhendi}.

  The basic design of the paper is as follows: In Section 2, we intend to give a short description of the field equations applicable to ECG and NLED. Eventually, we will deduce the simplest form of the independent field equation that specifies the topological black holes of ECG within the context of EN and LN forms of electrodynamics. In Section 3, we will be analyzing the physical characteristics of the resulting black holes and will compute the significant conserved and thermodynamic quantities. Subsequently, the final remarks are planned to be demonstrated in Section 4.

\section{Black holes in Einsteinian cubic gravity}
Einsteinian cubic gravity (ECG) serves as the most highly generic theory of gravity that remains independent of dimensions. It also incorporates a metric of spacetime along with the contractions of Riemann tensor. The linearized spectrum of this theory resembles with the one of ETG. Normally, we can illustrate the action that signifies ECG together with NLED as follows:
\begin{equation} \begin{split}
\mathcal{I}&=\frac{1}{16\pi}\int_{\mathcal{M}} d^4x\sqrt{-g}\bigg[R-2\Lambda_0+\sum_{j=2}^{3}\alpha_j\mathcal{L}_j-\lambda\mathcal{P}+\mathcal{L}_m(\mathcal{F})\bigg],
\label{1a}\end{split}
\end{equation}
where we select $c=G=1$. Additionally, $\Lambda_0$ serves as a bare cosmological constant, $\alpha_j$'s are the Lovelock coefficients and $\lambda$ is the cubic coupling parameter. Also, $\mathcal{L}_j$'s refers to $j$th-order Lovelock terms and we are adopting $\lambda\geq0$ everywhere in our work. Acknowledge that $\mathcal{L}_2$ is topological and $\mathcal{L}_3$ vanishes identically in four dimensions. One way to present the extra cubic contribution $\mathcal{P}$ is as \cite{4ar,1cubic}
\begin{equation}
	\begin{split}
		\mathcal{P}&=12R^{ab}_{cd}R^{ce}_{af}R^{df}_{be}+4R^{ab}_{cd}R^{ef}_{ab}R^{cd}_{ef}-12R^{ab}_{cd}R^c_aR^d_b+8R^a_bR^c_aR^b_c.\label{2a}
	\end{split}
\end{equation}
 Here, we will look into two representations of NLED, notably EN and LN models, whose corresponding Lagrangian densities are
 \begin{equation}\begin{split}
 		\mathcal{L}_{m}(\mathcal{F})=\left\{ \begin{array}{rcl}
 			 \beta^2\bigg(\exp{\big(-\frac{\mathcal{F}}{\beta^2}\big)}-1\bigg), &  & EN,\\-8\beta^2\ln{\bigg(1+\frac{\mathcal{F}}{8\beta^2}\bigg)}, & & LN,
 		\end{array}\right.\label{3a}
 	\end{split}
 \end{equation}
 wherein $\beta$ is the nonlinearity parameter and $\mathcal{F}=F_{\mu\nu}F^{\mu\nu}$ signifies the Maxwell's invariant with $F_{\mu\nu}=2\partial_{[\mu}A_{\nu]}$. Keep in mind that $A_{\nu}$ is the electromagnetic potential and the above Lagrangian densities are consistent with that of Maxwell's theory when $\beta$ is getting values closer to infinity. Even though the expressions specified by $\mathcal{L}_2$ and $\mathcal{L}_3$ are accordingly topological and trivial in four spacetime dimensions, the additional cubic term may nevertheless provide the dynamical implications to the field equations \cite{7ar}. To figure out the four dimensional black hole solution, we are adopting the following ansatz
\begin{equation}
	ds^2=-X^2(r)f(r)dt^2+\frac{dr^2}{f(r)}+r^2d\omega^2_{\kappa},\label{4a}
\end{equation}
where
\begin{equation}\begin{split}
		d\omega^2_{\kappa}=\left\{ \begin{array}{rcl}
			d\theta^2+\sin^2\theta d\phi^2, & 
			& \kappa=1, \\d\theta^2+\sinh^2\theta d\phi^2, &  & \kappa=-1,\\d\theta^2+d\phi^2, & & \kappa=0,
		\end{array}\right.\label{5a}
	\end{split}
\end{equation}
is the line element of a $2$-dimensional hyper-surface with a constant curvature $2\kappa$ and area $\Sigma_{\kappa}$. We also utilize the ansatz for electromagnetic potential as $A=h(r)dt$. One among the field equations that arise from varying Eq. (\ref{1a}) with regard to $f(r)$ is addressed by $X(r)=const$. Similarly, variation of action (\ref{1a}) relative to $h(r)$ gives rise to the nonlinear Maxwell's equation as follows:
\begin{equation}\begin{split}
		\left\{ \begin{array}{rcl}
			r^2\bigg(1+\frac{4}{\beta^2}\big(\frac{dh}{dr}\big)^2\bigg)\frac{d^2h}{dr^2}+2r\frac{dh}{dr}=0, &  & EN\\ r^2\bigg(1+\frac{1}{4\beta^2}\big(\frac{dh}{dr}\big)^2\bigg)\frac{d^2h}{dr^2}+2r\bigg(1-\frac{1}{4\beta^2}\big(\frac{dh}{dr}\big)^2\bigg)\frac{dh}{dr}=0,& & LN
		\end{array}\right..\label{6a}
	\end{split}
\end{equation}
 The aforementioned expressions are going to produce $h(r)$ as 
 \begin{equation}
 	\begin{split}
 	h(r)=\left\{ \begin{array}{rcl}
 		-\frac{1}{10}\big(4q^2\beta^2L_W\exp{(-L_W)}\big)^{\frac{1}{4}}\bigg[5+L_W\textbf{F}\bigg([1],\big[\frac{9}{4}\big],\frac{L_W}{4}\bigg)\bigg], &  & EN\\\frac{2q}{3r}\bigg[\frac{1}{1+\Gamma}-2\textbf{F}\bigg(\big[\frac{1}{4},\frac{1}{2}\big],\big[\frac{5}{4}\big],1-\Gamma^2\bigg)\bigg], & & LN
 	\end{array}\right.,\label{7a}	
 	\end{split}
 \end{equation}
 where $q$ is an integration constant that is directly related to the overall electric charge, $\Gamma=\sqrt{1+\frac{q^2}{\beta^2r^4}}$ and $L_W=LambertW(\frac{4q^2}{\beta^2r^4})$ such that $LambertW(x)\exp{\big(LambertW(x)\big)}=x$. 
 Eventually, the independent field equation for the metric function $f(r)$ will be offered by
\begin{equation}
	\begin{split}
		0=&-2r^2\bigg(\kappa-\Lambda_0r^2-r\big(\frac{df}{dr}\big)-f\bigg)-\frac{12\lambda}{r}\bigg[r^3f\bigg(\frac{d^2f}{dr^2}\bigg)^2+2\kappa r^2f\bigg(\frac{d^3f}{dr^3}\bigg)-4\kappa rf\bigg(\frac{d^2f}{dr^2}\bigg)\\&+r^3f\bigg(\frac{df}{dr}\bigg)\bigg(\frac{d^2f}{dr^2}\bigg)\bigg(\frac{d^3f}{dr^3}\bigg)^2+4\kappa f\bigg(\frac{df}{dr}\bigg)+4rf\bigg(\frac{df}{dr}\bigg)^2-\kappa r\bigg(\frac{df}{dr}\bigg)^2-4f^2\bigg(\frac{df}{dr}\bigg)\\&-2rf^2\bigg(r\bigg(\frac{d^3f}{dr^3}\bigg)-2\bigg(\frac{d^2f}{dr^2}\bigg)\bigg)\bigg]+\mathcal{N}_{\beta}, \label{8a}
	\end{split}
\end{equation}
where
\begin{equation}\begin{split}
		\mathcal{N}_{\beta}=\left\{ \begin{array}{rcl}
			-2r^4\beta^2+\frac{2r^2\beta q}{\sqrt{L_W}}(1-L_W), &  & EN\\-8r^4\beta^2+8r^4\beta^2\Gamma+8r^4\beta^2\ln{\big(\frac{2}{1+\Gamma}\big)}, & & LN
		\end{array}\right..\label{9a}
	\end{split}
\end{equation}
 Integration of Eq. (\ref{8a}) gives us
 \begin{equation}
 	\begin{split}
 		&\kappa r-m-\frac{\Lambda_0r^3}{3}-rf(r)+\mathcal{Y}(r)+\frac{\lambda}{r^2}\bigg[6rf(r)\bigg(\frac{d^2f}{dr^2}\bigg)\bigg(2\kappa+r\bigg(\frac{df}{dr}\bigg)-2f(r)\bigg)\\&-2r\bigg(\frac{df}{dr}\bigg)^2\bigg(3\kappa+r\bigg(\frac{df}{dr}\bigg)\bigg)-12f(r)\big(\kappa-f(r)\big)\bigg(\frac{df}{dr}\bigg)\bigg]=0, \label{10a}
 	\end{split}
 \end{equation}
 within which
 \begin{equation}
 	\begin{split}
 		\mathcal{Y}(r)=\left\{ \begin{array}{rcl}
 			\frac{q\beta r}{3\sqrt{L_W}}\bigg[1+L_W+\frac{4}{5}L_W^2\textbf{F}\bigg([1],\big[\frac{9}{4}\big],\frac{L_W}{4}\bigg)\bigg], &  & EN\\\frac{16q^2}{9r}\textbf{F}\bigg(\big[\frac{1}{2},\frac{1}{4}\big],\big[\frac{5}{4}\big],1-\Gamma^2\bigg)+\frac{4\beta^2r^3}{9}\bigg[3\ln{\bigg(\frac{1+\Gamma}{2}\bigg)+5(1-\Gamma)}\bigg], & & LN
 		\end{array}\right..\label{11a}	
 	\end{split}
 \end{equation}
  Here, $m$ is an integration constant that is linked to the entire mass of black hole. The solution of ETG sourced by Born-Infeld-type electromagnetism can easily be deduced if one puts $\lambda=0$ in Eq. (\ref{11a}), so we can write
   \begin{eqnarray}\begin{split}
   		f(r)&=\kappa-\frac{m}{r}-\frac{\Lambda_0r^2}{3}\\&+\left\{ \begin{array}{rcl}
   			\frac{q\beta }{3\sqrt{L_W}}\bigg[1+L_W+\frac{4}{5}L_W^2\textbf{F}\bigg([1],\big[\frac{9}{4}\big],\frac{L_W}{4}\bigg)\bigg], &  & EN\\\frac{16q^2}{9r^2}\textbf{F}\bigg(\big[\frac{1}{2},\frac{1}{4}\big],\big[\frac{5}{4}\big],1-\Gamma^2\bigg)+\frac{4\beta^2r^2}{9}\bigg[3\ln{\bigg(\frac{1+\Gamma}{2}\bigg)+5(1-\Gamma)}\bigg], & & LN
   		\end{array}\right.,\label{12a}\end{split}
   \end{eqnarray} 
   where the total mass per unit area associated with the above Eq. (\ref{12a}) in ETG is given by \cite{56ar,57ar,58ar,59ar,60ar} 
   \begin{eqnarray}
   	M_{ETG}=\frac{m}{8\pi}.\label{13a}
   \end{eqnarray}
   
 \section{Thermodynamic quantities, first law and thermodynamic stability}
 In this part, we will explore the thermodynamic aspects of the black holes characterized by Eq. (\ref{10a}). For this purpose, we first have to check the response of this equation near the black hole horizon $r_+$. The Taylor series representation of metric function in the vicinity of event horizon is given by
\begin{equation}
	f(r)=\sum_{n=0}^{\infty}c_n(r-r_+)^n,\label{14a}
\end{equation}
where $c_n=\frac{f^{(n)}(r_+)}{n!}$. Note that $c_0=0$, via condition $f(r_+)=0$ and $c_1=f'(r_+)=2\kappa_g$. Here, $\kappa_g$ refers to the surface gravity on the event horizon and $f'(r_+)$ is non-negative everywhere. Hence, by utilizing the above series expansion in Eq. (\ref{10a}), we obtain the expression up to second order of $(r-r_+)$ as follows:
\begin{equation}\begin{split}
&\kappa r_+-m-8\lambda\kappa_g^2\big(2\kappa_g+\frac{3\kappa}{r_+}\big)-\frac{\Lambda_0r_+^3}{3}+\mathcal{Y}(r_+)+\bigg[\kappa-\Lambda_0r_+^2-2\kappa_gr_+\\&-\frac{24\lambda\kappa\kappa_g^2}{r_+^2}+\mathcal{Y}'(r_+)\bigg](r-r_+)+\bigg[72c_3\lambda\kappa_g^2+\frac{72c_3\lambda\kappa\kappa_g}{r_+}+24c_2^2\lambda\kappa_g-c_2r_+\\&-2\kappa_g-\Lambda_0r_+-\frac{96c_2\lambda\kappa_g^2}{r_+}-\frac{72c_2\lambda\kappa\kappa_g}{r_+^2}+\frac{72\lambda\kappa\kappa_g^2}{r_+^3}+\frac{96\lambda\kappa_g^3}{r_+^2}+\frac{1}{2}\mathcal{Y}''(r_+)\bigg](r-r_+)^2\\&+O((r-r_+)^3)=0,\label{15a}\end{split}
\end{equation}
where
\begin{equation}
	\begin{split}
		\mathcal{Y}(r_+)=\left\{ \begin{array}{rcl}
			\frac{q\beta r_+}{3\sqrt{L_W(r_+)}}\bigg[1+L_W(r_+)+\frac{4}{5}L_W^2(r_+)\textbf{F}\bigg([1],\big[\frac{9}{4}\big],\frac{L_W(r_+)}{4}\bigg)\bigg], &  & EN\\\frac{16q^2}{9r_+}\textbf{F}\bigg(\big[\frac{1}{2},\frac{1}{4}\big],\big[\frac{5}{4}\big],1-\Gamma_+^2\bigg)+\frac{4\beta^2r_+^3}{9}\bigg[3\ln{\bigg(\frac{1+\Gamma_+}{2}\bigg)+5(1-\Gamma_+)}\bigg], & & LN
		\end{array}\right..\label{16a}	
	\end{split}
\end{equation}
Here, $\Gamma_+=\sqrt{1+\frac{q^2}{\beta^2r_+^4}}$ and $L_W(r_+)=LambertW(\frac{4q^2}{\beta^2r_+^4})$. Furthermore, 
\begin{equation}
	\begin{split}
		\mathcal{Y}'(r_+)=\left\{ \begin{array}{rcl}
			-\frac{q\beta (L_W(r_+)-1)}{\sqrt{L_W(r_+)}}, &  & EN\\\frac{4r_+^2\beta^2}{9}\bigg[27-27\Gamma_++\frac{8q^2}{r_+^4\beta^2\Gamma_+}+27\ln{(\frac{1+\Gamma_+}{2})}\bigg], & & LN
		\end{array}\right.,\label{17a}	
	\end{split}
\end{equation}
and 
\begin{equation}
	\begin{split}
		\mathcal{Y}''(r_+)=\left\{ \begin{array}{rcl}
			\frac{2q\beta }{r_+\sqrt{L_W(r_+)}}, &  & EN\\-\frac{64q^2\beta^4r_+^5}{9\beta^4r_+^8\Gamma_+^3}+24r_+\beta^2\ln{(\frac{1+\Gamma_+}{2})}, & & LN
		\end{array}\right..\label{18a}	
	\end{split}
\end{equation}
From Eq. (\ref{15a}), it is possible to obtain
\begin{equation}
	\kappa r_+-m-8\lambda\kappa_g^2\big(2\kappa_g+\frac{3\kappa}{r_+}\big)-\frac{\Lambda_0r_+^3}{3}+\mathcal{Y}(r_+)=0,\label{19a}
\end{equation}
and
\begin{equation}
	\kappa-\Lambda_0r_+^2-2\kappa_gr_+-\frac{24\lambda\kappa\kappa_g^2}{r_+^2}+\mathcal{Y}'(r_+)=0.\label{20a}
\end{equation}
Using the limit $\beta\rightarrow\infty$, the above equations are producing the results corresponding to ECG with Maxwell's source \cite{7ar}
\begin{equation}
	\kappa r_+-m-8\lambda\kappa_g^2\big(2\kappa_g+\frac{3\kappa}{r_+}\big)-\frac{\Lambda_0r_+^3}{3}+\frac{q^2}{4r_+}=0,\label{21a}
\end{equation}
and
\begin{equation}
	\kappa-\Lambda_0r_+^2-2\kappa_gr_+-\frac{24\lambda\kappa\kappa_g^2}{r_+^2}-\frac{q^2}{4r_+^2}=0.\label{22a}
\end{equation}
 This behavior is similar to the situation of black hole solution of ECG that is supported by the Born-Infeld electromagnetic field \cite{1cubic}. Additionally, the resulting black hole solutions governed by Eq. (\ref{10a}) are possessing regular horizons and the regularity criteria will generate the one-parameter family of solutions instead of two-parameter family \cite{1cubic}. Utilizing Eqs. (\ref{19a}) and (\ref{20a}), it is simple to obtain the expressions of parameter $m$ and surface gravity as follows:
 \begin{equation}
 	\begin{split}
 		m=\kappa r_+-16\lambda\kappa_g^3-\frac{24\kappa\lambda\kappa_g^2}{r_+}-\frac{\Lambda_0r_+^3}{3}+\mathcal{Y}(r_+),\label{23a}
 	\end{split}
 \end{equation}
 \begin{equation}
 	\begin{split}
 		\kappa_g=-\frac{r_+^3}{24\lambda\kappa}+\frac{r_+^3}{24\lambda\kappa}\sqrt{\bigg[1-\frac{24\Lambda_0\kappa\lambda}{r_+^2}+\frac{24\lambda\kappa^2}{r_+^4}+\frac{24\lambda\kappa}{r_+^4}\mathcal{Y}'(r_+)\bigg]}.\label{24a}
 	\end{split}
 \end{equation}
    The basic idea of surface gravity is capable of being employed to illustrate the Hawking temperature in the form \cite{Hawk}
\begin{equation}
	T_H=\frac{\kappa_g}{2\pi},\label{25a}
\end{equation}
wherein $\kappa_g$ is expressed in Eq. (\ref{24a}). 
Additionally, $m$ in Eq. (\ref{23a}) has connections with the total mass $M$ of black hole via
\begin{equation}
	M=\bigg(1-\frac{8\lambda\Lambda^2}{3}\bigg)\frac{m}{8\pi}.\label{26a}
\end{equation} 
If the conception of ECG is unitary and ghost-free, it would turnout essential desirable to proceed with $\bigg(1-\frac{8\lambda\Lambda^2}{3}\bigg)>0$ \cite{61ar,62ar,63ar}. Point out that $\Lambda$ is the effective cosmological constant and is fulfilling the algebraic equation \cite{1cubic}
\begin{equation}
	\frac{8\lambda}{9}\Lambda^3-\Lambda+\Lambda_0=0.\label{27a}
\end{equation}
 The discriminant for this specific Eq. (\ref{27a}) serves as
 \begin{equation}
 	\Delta=\frac{32\lambda}{9}\big(1-6\lambda\Lambda_0^2\big).\label{28a}
 \end{equation}
 In consequence, whenever $\Delta\geq0$ or $\lambda\big(1-6\lambda\Lambda_0^2\big)\geq0$, Eq. (\ref{27a}) will feature three real roots . One of various possibilities is $\Lambda_0=0$, which can result in $\Lambda=0$ and $\Lambda=\pm\frac{3}{2\sqrt{2\lambda}}$. If we prefer $\Lambda=\pm\frac{3}{2\sqrt{2\lambda}}$, then $\bigg(1-\frac{8\lambda\Lambda^2}{3}\bigg)$ becomes negative, which goes against unitarity. Alternatively, the diminishing of $\Lambda_0$ gives rise to the unitary model when our resulting solution is asymptotically flat, i.e. when $\Lambda=0$. Another possibility is $\Lambda_0^2=1/6\lambda$ which is analogous to $\Delta=0$. It is relatively simple to make sure that for all the values of $\Lambda$ relevant to this situation, the respective model is no further unitary simply because the term $\bigg(1-\frac{8\lambda\Lambda^2}{3}\bigg)$ is becoming non-positive. For several other values of parameters including those that line up with $\Delta<0$, it might be possible to extract some vacuum solutions within a unitary model. In this specific instance, the effective cosmological constant in Eq. (\ref{27a}) is exhibiting one real and two complex conjugate values \cite{1cubic}. Here, we are emphasizing on the scenario where $\Lambda_0$ diminishes. Thereby, only solutions with $\Lambda=0$ are acceptable. So, the mass per unit area can be specified as
 \begin{equation}
 	\begin{split}
 	M=\frac{m}{8\pi}=\frac{1}{8\pi}\bigg[\kappa r_+-16\lambda\kappa_g^3-\frac{24\kappa\lambda\kappa_g^2}{r_+}+\mathcal{Y}(r_+)\bigg].\label{29a}
 	\end{split}
 \end{equation}
  At the same time, $T_H$ might be presented as
\begin{equation}\begin{split}
T_H(r_+)&=-\frac{r_+^3}{48\pi\lambda\kappa}+\frac{r_+^3}{48\pi\lambda\kappa}\sqrt{\bigg[1+\frac{24\lambda\kappa^2}{r_+^4}+\frac{24\lambda\kappa}{r_+^4}\mathcal{Y}'(r_+)\bigg]}.\label{30a}\end{split}
\end{equation}
\begin{figure}[h]
	\centering
	\includegraphics[width=0.8\textwidth]{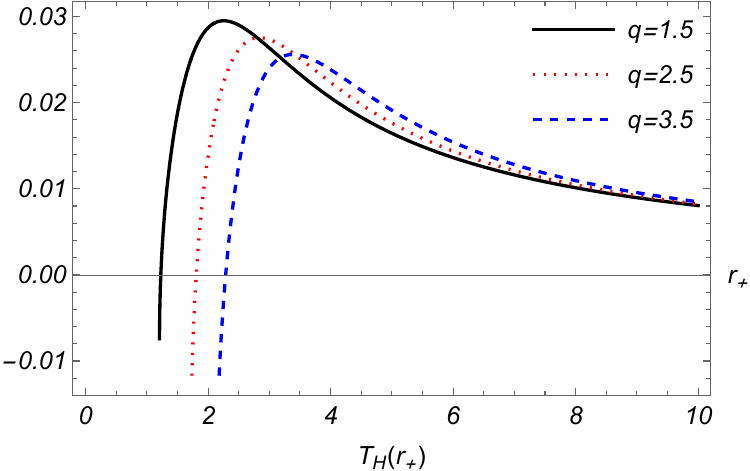}
	\caption{Evolution of $T_H(r_+)$ (Eq. (\ref{30a})) for numerous values of $q$. We have taken $\kappa=1$, $\beta=0.5$ and $\lambda=1.5$.}\label{fatih6}
\end{figure}
\begin{figure}[h]
	\centering
	\includegraphics[width=0.8\textwidth]{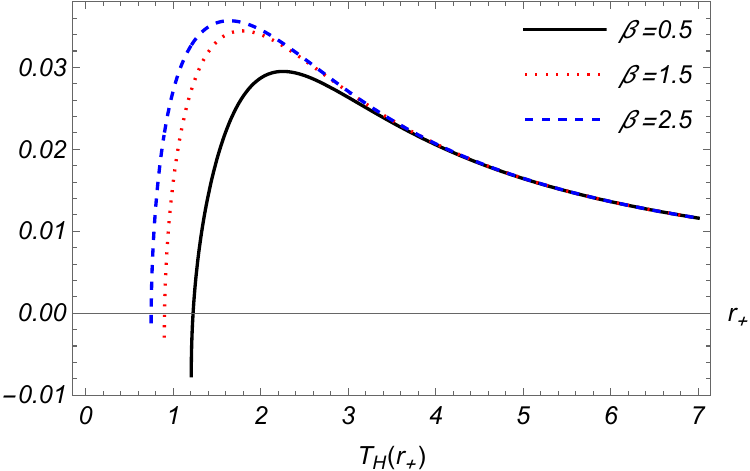}
	\caption{Evolution of $T_H(r_+)$ (Eq. (\ref{30a})) with several choices of $\beta$. We have taken $\kappa=1$, $q=1.5$ and $\lambda=1.5$.}\label{fatih7}
\end{figure}
\begin{figure}[h]
	\centering
	\includegraphics[width=0.8\textwidth]{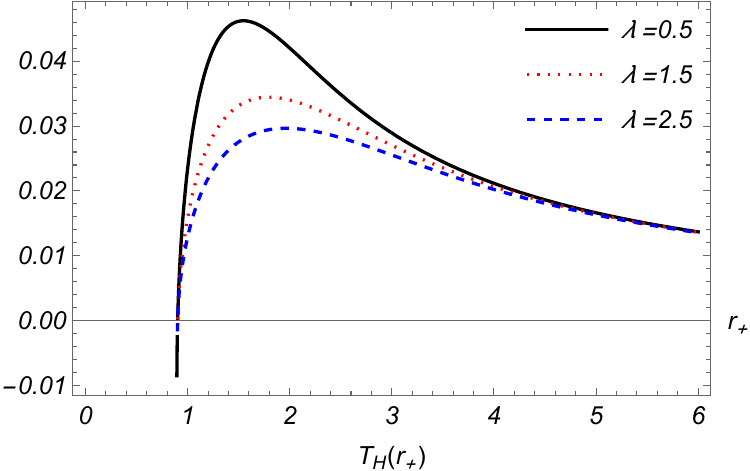}
	\caption{Evolution of $T_H(r_+)$ (Eq. (\ref{30a})) with several choices of $\lambda$. We have taken $\kappa=1$, $q=1.5$ and $\beta=1.5$.}\label{fatih8}
\end{figure}

Figs. \ref{fatih6}-\ref{fatih8} outline how electric charge, nonlinear electric field and cubic coupling parameter are altering the Hawking temperature (\ref{30a}) of resulting black hole. Note that we primarily examine the effects of LN electromagnetic field on the thermodynamic characteristics of our finalized solution throughout our work since we feel that the plots of thermal quantities within the backdrop of EN electrodynamics are roughly comparable to those in LN framework. The black hole would be regarded physical if $T_H$ appears positive. The event horizon of an extremal black hole is encountered at location $r_1$ relating to which $T_H$ is zero. It is obvious to notice that this specific value elevates with the rise in $q$ and falls back when $\beta$ is growing. It should also be pointed out that the cubic parameter encounters zero impact on $r_1$. Plus, $T_H$ firstly advances with reference to $r_+$, which signifies that smaller black hole experiences more thermal sensitivity as its relative size gets larger. For any value of electric charge, nonlinearity parameter and cubic coupling constant, the temperature escalates to its peak and then slowly declines when $r_+$ is going up. In addition, the optimum value of $T_H$ is getting lower when $q$ and $\lambda$ are elevated. By contrast, the maximum temperature begins to rise when $\beta$ goes higher. These figures also suggest that the parameters within the current setup endure a significant impact on the temperature of smaller black holes, whilst their impressions on the temperature of larger objects are inconsequential.

The overall electric charge of our simulated black hole solutions can be ascertained by projecting the flux of electric field at infinity via Gauss law as
\begin{equation}
	\begin{split}
		Q=\frac{1}{4\pi}\int r^2\frac{\partial\mathcal{L}_m}{\partial\mathcal{F}}F_{\alpha\beta}n^{\alpha}u^{\beta}d\omega_{\kappa},\label{31a}
	\end{split}
\end{equation}
 within which $\mathcal{L}_m$ is the Lagrangian density presented in Eq. (\ref{3a}). Furthermore, $n^{\alpha}$ and $u^{\beta}$ stand for the spacelike and timelike unit normals, respectively. Thus, by using the ansatz (\ref{4a}) and the Lagrangian densities of both EN and LN models, we come up with
 \begin{equation}
 	Q=\frac{q}{16\pi}.\label{32a}
 \end{equation}
  This signifies that the overall electric charge has no dependence on $\beta$. Since Eq. (\ref{10a}) indicates the charged black holes in ECG, thereby rendering the area law is unreliable in assessing the entropy \cite{92,93}. In consequence, Hamiltonian approach is capable of being employed to quantify this quantity \cite{94,95,96}. This conventional technique was additionally implemented for the black holes of higher curvature theories, where Wald's formula is carried out to ascertain the black hole's entropy \cite{97,98}. Such a formula is often stated as
  \begin{equation}
  	S=-2\pi\int_{H}d^2x\sqrt{\gamma}\frac{\delta\mathcal{L}_g}{\delta R_{\alpha\beta\rho\sigma}}\epsilon_{\alpha\beta}\epsilon_{\rho\sigma},\label{33a}
  \end{equation}
  where $\gamma$ serves as the determinant of stimulated metric $\gamma_{\alpha\beta}$ that identifies the horizon's geometry. Likewise, $\frac{\delta\mathcal{L}_g}{\delta R_{\alpha\beta\rho\sigma}}$ points to the Euler-Lagrange derivative of gravitational Lagrangian, and $\epsilon_{\alpha\beta}$ indicates the binormal to the horizon and meets the normalizing condition $\epsilon_{\alpha\beta}\epsilon^{\alpha\beta}=-2$.  By executing the Wald's formula on our model, i.e. action (\ref{1a}), we may infer
  \begin{equation}
  	\begin{split}
  		S&=\frac{1}{4}\int_{H}d^2x\sqrt{\gamma}\bigg[1+2\alpha_2R_{H}+\lambda\bigg(36R^{\mu\nu}_{\beta\sigma}R_{\alpha\mu\rho\nu}+3R^{\mu\nu}_{\alpha\beta}R_{\rho\sigma\mu\nu}-12R_{\alpha\rho}R_{\sigma\beta}\\&-24R^{\mu\nu}R_{\mu\beta\nu\rho}g_{\alpha\sigma}+24g_{\beta\sigma}R_{\rho\mu}R^{\mu}_{\alpha}\bigg)\epsilon^{\alpha\beta}\epsilon^{\rho\sigma}\bigg],\label{34a}
  	\end{split}
  \end{equation}
  in which $R_{H}$ is symbolizing the Ricci scalar connected to the metric $\gamma_{\alpha\beta}$. The main reason behind the appearance of this term is the involvement of the Gauss-Bonnet Lagrangian $\mathcal{L}_2=R^2-4R_{\alpha\beta}R^{\alpha\beta}+R_{\alpha\beta\rho\sigma}R^{\alpha\beta\rho\sigma}$in Eq. (\ref{1a}). Even though it is topological in four dimensions and is not unable to form part of Eq. (\ref{10a}), however, it shows up in the entropy's expression. As such, by implementing the metric (\ref{4a}) with $X(r)=1$, we can quantify $S$ as follows:
\begin{equation}\begin{split}
		S&=\frac{r_+^2}{4}\bigg[1-\frac{24\lambda\kappa_g^2}{r_+^2}\bigg(1+\frac{2\kappa}{\kappa_gr_+}\bigg)\bigg]+\kappa\alpha_2.\label{35a}\end{split}
\end{equation}
 Observe that $\kappa_g$ is specified in Eq. (\ref{24a}). The electric potential of the event horizon estimated at infinity might be quantified as \cite{98a,98b}
 \begin{equation}\begin{split}
 	\psi(r_+)&=A_{\alpha}\mathcal{Z}^{\alpha}\big|_{r\rightarrow\infty}-A_{\alpha}\mathcal{Z}^{\alpha}\big|_{r=r_+},\label{36a}\end{split}
 \end{equation}
  where $\mathcal{Z}^{\alpha}=\partial/\partial t$ serves as the event horizon's null generator. Accordingly, utilizing the foregoing expression (\ref{36a}) and Eq. (\ref{7a}), we identify
  \begin{equation}
  	\begin{split}
  		\psi(r_+)=\left\{ \begin{array}{rcl}
  			\frac{1}{10}\big(4\beta^2L_W(r_+)\exp{(-L_W(r_+))}\big)^{\frac{1}{4}}\sqrt{16\pi Q}\bigg[5+L_W(r_+)\textbf{F}\bigg([1],\big[\frac{9}{4}\big],\frac{L_W(r_+)}{4}\bigg)\bigg], &  & EN\\-\frac{32\pi Q}{3r_+}\bigg[\frac{1}{1+\Gamma_+}-2\textbf{F}\bigg(\big[\frac{1}{4},\frac{1}{2}\big],\big[\frac{5}{4}\big],1-\Gamma_+^2\bigg)\bigg], & & LN
  		\end{array}\right..\label{37a}	
  	\end{split}
  \end{equation}
   In order to inspect the validity of first law, we should need to present a Smarr-type formula. Hence, upon the utilization of Eqs. (\ref{29a}) and (\ref{32a}), we might express it as
   \begin{equation}
   	\begin{split}
   		M=\frac{1}{8\pi}\bigg[\kappa r_+-16\lambda\kappa_g^3-\frac{24\kappa\lambda\kappa_g^2}{r_+}+\mathcal{Y}(r_+)\bigg],\label{38a}
   	\end{split}
   \end{equation}
   within which $\mathcal{Y}(r_+)$ is defined in Eq. (\ref{16a}), $\Gamma_+=\sqrt{1+\frac{(16\pi Q)^2}{\beta^2r_+^4}}$ and $L_W(r_+)=LambertW(\frac{4(16\pi Q)^2}{\beta^2r_+^4})$.
   From entropy (\ref{35a}), we can presume that $r_+=r_+(S,Q)$ and most generally the finite mass $M$, as portrayed in Eq. (\ref{38a}), is also believed to rely on entropy $S$ and electric charge $Q$. In other words, we might say that both $S$ and $Q$ are the extensive variables for $M$. Thereby, temperature (\ref{30a}) and potential (\ref{37a}) are believed to be treated like conjugate intensive variables associated with entropy and electric charge, respectively. So, one determines
   \begin{equation}
   	T_H=\bigg(\frac{\partial M}{\partial S}\bigg)_Q=\bigg(\frac{\partial M}{\partial r_+}\bigg)_Q\bigg(\frac{\partial S}{\partial r_+}\bigg)^{-1}_Q,\label{39a}
   \end{equation}
    and 
    \begin{equation}
    	\psi=\bigg(\frac{\partial M}{\partial Q}\bigg)_{S}=\bigg(\frac{\partial M}{\partial Q}\bigg)_{r_+}-T_H\bigg(\frac{\partial S}{\partial Q}\bigg)_{r_+}.\label{40a}
    \end{equation}
   The results of our analysis verify that $T_H$ and $\psi$ specified respectively in Eqs. (\ref{39a}) and (\ref{40a}) correlate with the respective expressions reported in Eqs. (\ref{30a}) and (\ref{37a}) for charged black hole with $\Lambda_0=0$ and $\kappa=1$. Accordingly, the first law of black hole thermodynamics \cite{99,100,101,102,103} could possibly be carried out as
\begin{equation}
	dM=T_HdS+\psi dQ. \label{41a} 
\end{equation} 

The formula provided for heat capacity is usually acknowledged as 
\begin{equation}
	C_h=T_{H}\frac{dS}{dT_H}\bigg|_{Q}. \label{42a}
\end{equation}
\begin{figure}[h]
	\centering
	\includegraphics[width=0.8\textwidth]{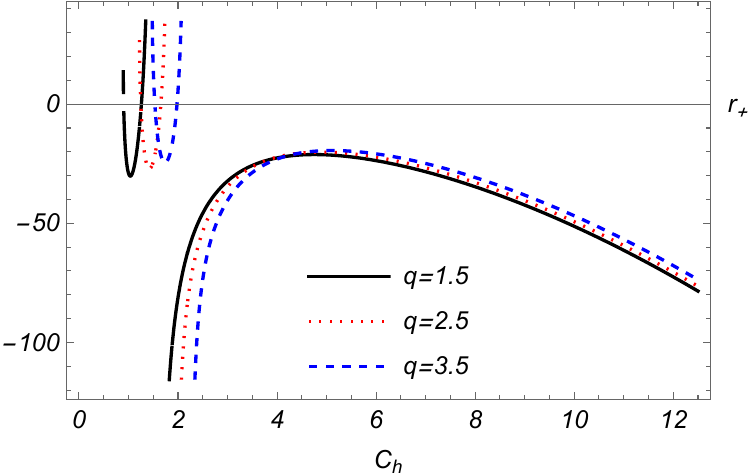}
	\caption{Progress of the heat capacity with different amounts of $q$. Our further presumptions are $\kappa=1$, $\lambda=0.5$ and $\beta=1.5$.}\label{fatih10}
\end{figure}
\begin{figure}[h]
	\centering
	\includegraphics[width=0.8\textwidth]{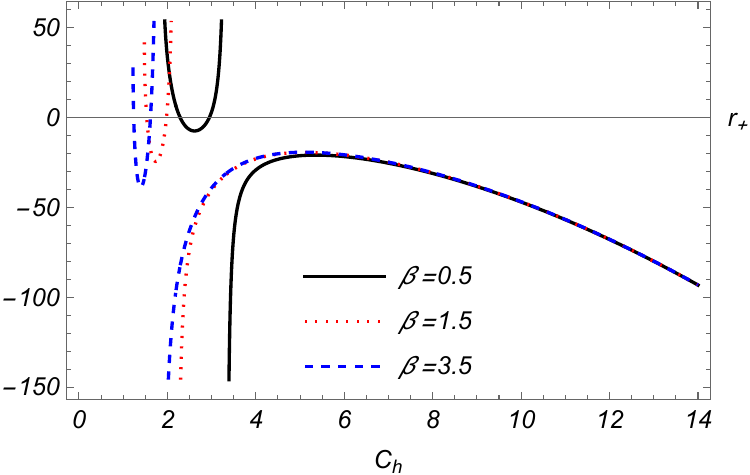}
	\caption{Progress of the heat capacity with several levels of $\beta$. Our further presumptions are $\kappa=1$, $\lambda=0.5$ and $\beta=3.5$.}\label{fatih11}
\end{figure}
\begin{figure}[h]
	\centering
	\includegraphics[width=0.8\textwidth]{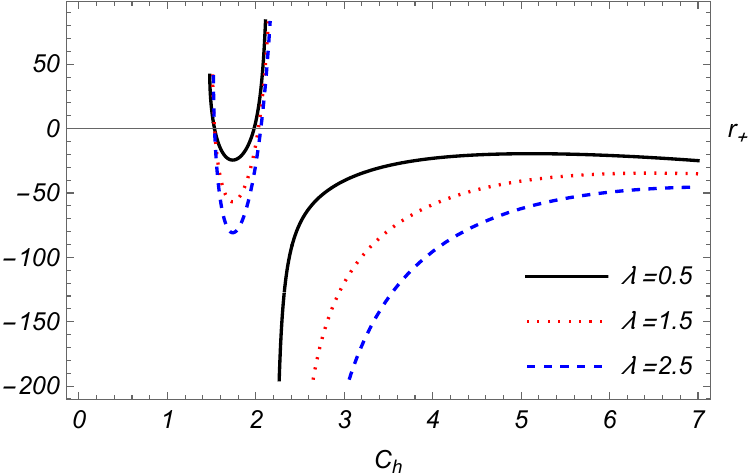}
	\caption{Progress of the heat capacity (Eq. (\ref{42a})) for multiple levels of $\lambda$. Our further presumptions are $\kappa=1$, $q=3.5$ and $\beta=1.5$.}\label{fatih12}
\end{figure}

Figs. \ref{fatih10}-\ref{fatih12} disclose the manner in which $C_h$ is manipulated by  $q$, $\beta$ and $\lambda$. Local stability in canonical ensemble is ascertained provided $T_H$ and $C_H$ are both positive. The horizon radius whereby $C_h$ is declared ill-defined is recognizing the possible emergence of a second-order phase transition, whilst the point where it vanishes recommends the prospect of a first-order transition. We assessed that heat capacity has two zeros, which are positioned at $r_+=r_1$ and at $r_+=r_2$. This indicates that the black hole becomes fully unstable when either $r_+$ is linked to $(0,r_1)$ or happens to fall in $(r_1,r_2)$. This happens since, neither $T_H$ nor $C_h$ are positive in these intervals. We identify that the object is solely locally stable when $r_+$ fits within $(r_2,r_3)$. At $r_+=r_3$, heat capacity has emerged as irregular, and the black hole truly again deemed unstable for whatever value of $r_+$ that drops in $(r_3,\infty)$. We witnessed that as the magnitudes of $q$ and $\lambda$ are advancing, both the horizon radii $r_2$ and $r_3$ become bigger. Interestingly, when $\beta$ approaches higher magnitudes, the horizon radius $r_2$ gets bigger, whereas $r_3$ which reflects to the position infinite $C_h$ is diminishing. It must be emphasized that the horizon radius $r_1$ is correlating to the extremal black hole's horizon radius whereby temperature and heat capacity vanish. In grand canonical ensemble, charge $Q$ must be acknowledged for being a thermodynamic entity alongside entropy. Along with $T_h>0$ and $C_h>0$, local stability seems affirmed when the determinant of Hessian matrix $detH^M$ is positive \cite{97,98}. The Hessian matrix is occasionally quantified as 
\begin{equation}
	\textbf{H}=
	\left[ {\begin{array}{ccc}
			\frac{\partial^2M}{\partial S^2} & \frac{\partial^2M}{\partial S\partial Q} \\
			\frac{\partial^2M}{\partial Q\partial S} & \frac{\partial^2M}{\partial Q^2}\\
	\end{array} } \right],\label{A48}
\end{equation}
whereby $\frac{\partial^2M}{\partial S\partial Q}=\frac{\partial^2M}{\partial Q\partial S}$. The entries of the previously described matrix might be estimated through the utilization of entire mass (\ref{29a}), Hawking temperature (\ref{30a}) and entropy (\ref{35a}). Since $C_h$ and $detH^M$ owned extremely sophisticated precise forms, we are avoiding to outline their expressions.
\begin{figure}[h]
	\centering
	\includegraphics[width=0.8\textwidth]{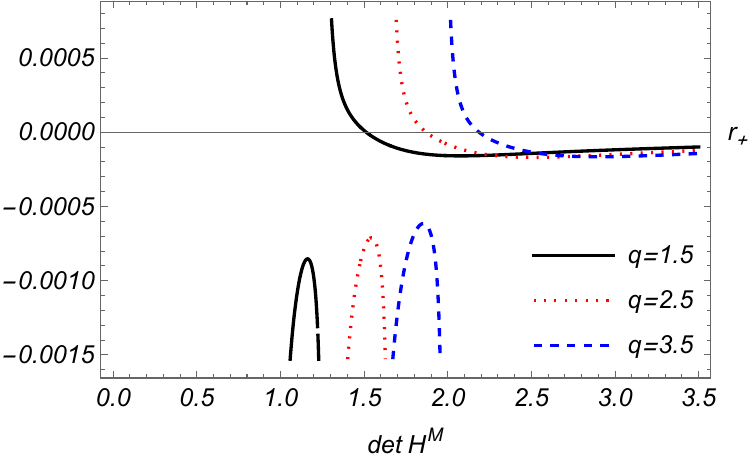}
	\caption{Plot of $detH^M$ for multiple values of $Q=\frac{q}{16\pi}$. We are taking $\beta=1.5$, $\kappa=1$ and $\lambda=0.5$.}\label{fatih15}
\end{figure}
\begin{figure}[h]
	\centering
	\includegraphics[width=0.8\textwidth]{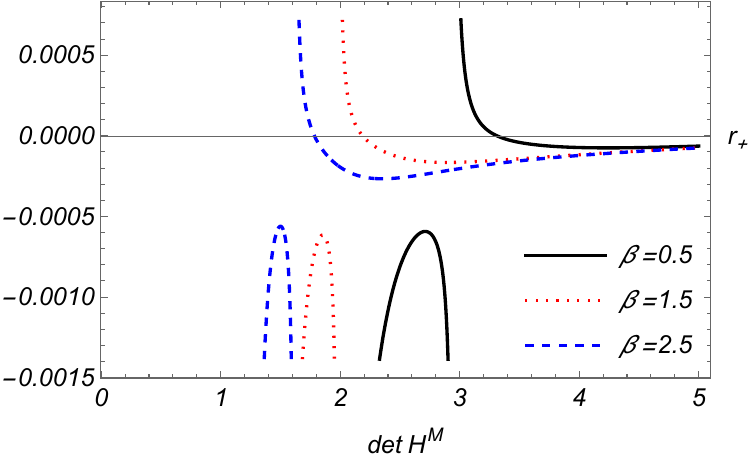}
	\caption{Plot of $det H^M$ for multiple values of $\beta$. We are taking $q=3.5$, $Q=\frac{q}{16\pi}$, $\kappa=1$ and $\lambda=0.5$.}\label{fatih16}
\end{figure}
\begin{figure}[h]
	\centering
	\includegraphics[width=0.8\textwidth]{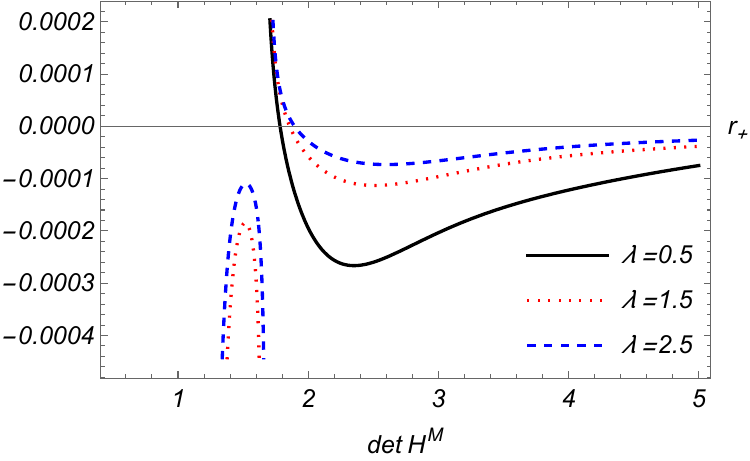}
	\caption{Plot of $detH^M$ for multiple values of $\lambda$. We are taking $q=3.5$, $Q=\frac{q}{16\pi}$, $\kappa=1$ and $\beta=2.5$}\label{fatih17}
\end{figure} 

Figs. \ref{fatih15}-\ref{fatih17} reveal the characterization of $detH^M$ as an expression of $r_+$. We observed that $detH^M$ has turned undefinable at $r_+=r_a$, whilst vanishes at $r_+=r_b$, recommending that the object with $T_H>0$ and $C_h>0$ shows up to act as locally stable if its respective $r_+$ conforms to a criteria requiring $r_a<r_+<r_b$. In contrast, any object with $r_+$ falls in either $(0,r_a)$ or in $(r_b,\infty)$ happens to be unstable in the present ensemble simply because the respective $detH^M$ is deemed negative within these ranges. We noticed that both the particular values $r_a$ and $r_b$ are acquiring higher magnitudes when the electric charge grows. One may analyze the opposite responses from these horizon radii as both are shrinking with the rise in $\beta$. Meanwhile, the value $r_a$ remains unaffected, while $r_b$ rises as the cubic coupling parameter raises.

\section{Conclusion and Outlook}

Here, we conduct the investigation on four-dimensional charged black holes of ECG. We relied on the models of EN and LN electromagnetic fields to identify the matter sources of ECG. In this backdrop, we first presented the action function and derived the respective differential equations of motion for the description of ECG minimally coupled to EN and LN electromagnetic theories. To identify an innovative category of charged topological black holes, we have utilized the nonlinear Maxwell's equations for figuring out the forms of electromagnetic potential within both EN and LN electromagnetic frameworks. We determined a second order differential equation for the metric function, which relies upon the topological parameter, geometric mass, coupling parameter of ECG, electric charge and nonlinearity parameter of NLED. We realized that the derived equation goes back to the respective solutions of ETG when $\lambda$ vanishes. To maintain the model unitary and rid of ghosts, we inquired into the thermodynamic and physical features of black holes for the scenario of vanishing $\Lambda_0$. The identified solution within this situation was correlated with the asymptotically flat objects that carrying spherical topology. Beyond that, we estimated the appropriate thermodynamic quantities and focused on the local stability in both canonical and grand canonical ensembles. We arrived at the formulas of $M$ and $T_H$ by incorporating the power series expansion of $f(r)$ centered at the event horizon up to second order in the equations of motion (\ref{10a}). We also communicated about how the progress of $T_H$ is dictated by electric charge $q$, nonlinearity parameter $\beta$ and cubic coupling parameter $\lambda$. We reported that the maximum value of $T_H$ drops with improvement in $q$ and $\lambda$, whereas it is raising whenever $\beta$ is elevated. We additionally noticed that as $\beta\rightarrow\infty$, the outcomes of Maxwell's theory from both the EN and LN models are retrieved. Moreover, we found that the event horizon of extreme black hole $r_1$ grows with increase in $q$ and shrinks with increase in $\beta$. Conversely, it becomes revealed that $\lambda$ has no significant impact on the extremal value $r_1$. We assessed the forms of electric potential in both scenarios and implemented Wald's formula for quantifying the entropy. We identified that the thermodynamic first law meets the specifications of our finalized black holes in ECG within both frameworks.

 We looked into how $q$, $\beta$ and $\lambda$ are dictating the progressions of $T_H$ and $C_H$ in four dimensions. The subsequently described charged black holes in ECG with EN and LN electromagnetic sources specified by Eq. (\ref{10a}) are supposed to be thermally stable in canonical ensemble when both $T_H$ and $C_h$ are positive. Against this backdrop, the values of $r_+$ that sustain a concordance with either $T_H<0$ or $C_h<0$ are identifying the region of thermal instability. In accordance with the evolution of $C_h$, we also inferred that heat capacity has two zeros, i.e. $r_1$ and $r_2$. Likewise, it has also a singularity $r_3$, i.e. a certain value of $r_+$ at which it becomes irregular. Consequently, the object will merely be unstable in $(0,r_1)\cup(r_1,r_2)$ as both $T_H$ and $C_h$ are not positive within this range. The object is declared stable only as long as the respective event horizon drops in $(r_2,r_3)$. Similarly, the object is rendered unstable when $r_+$ is within $(r_3,\infty)$. Figs. \ref{fatih10}-\ref{fatih12} also revealed that the particular horizon radii $r_2$ and $r_3$ are getting bigger with higher electric charge and cubic coupling parameter. In addition, $r_2$ will grow and $r_3$ shrinks when $\beta$ gets larger values. Furthermore, the zero $r_1$ of $C_h$ is identical to the extremal horizon radius. Accordingly, any black hole which satisfies $r_+<r_1$ is not truly physical. The positive values of $detH^M$, $T_H$ and $C_h$ are indicative of local stability within the grand canonical ensemble. Figs. \ref{fatih15}-\ref{fatih17} demonstrated that the objects have become stable in this ensemble when their respective $r_+$'s are within interval $(r_a,r_b)$. Outside of this range, the black hole would be experiencing instability. Point out that $detH^M$ is ill defined at $r_+=r_a$, while vanishes at $r_+=r_b$. We also spotted that these horizon values are going up when electric charge is rising. Contrarily, these horizon radii are declining when $\beta$ gets higher values. At last, we observed that when $\lambda$ is on the rise, $r_a$ is not effected, whereas $r_b$ is growing.
 
 Investigating the repercussions of Chaplygin-like dark fluid, cloud of strings, and quintessential dark energy on the physical aspects of black holes in ECG would likely be of great significance. Beyond that, the thermodynamic attributes of black holes in ECG coupled to Yang-Mills theories could possibly be quite intriguing. We have set aside these strategies for subsequent work.

\end{document}